\documentclass[aps,prl,twocolumn,nofootinbib]{revtex4-1}

\usepackage{graphicx}

\begin{document}

\title{Perturbative $\lambda$-Supersymmetry and Small $\kappa$-Phenomenology}

\author{Sibo Zheng}
\affiliation{Department of Physics, Chongqing University, Chongqing 401331, P. R. China}

\begin{abstract}
For the minimal $\lambda$-supersymmetry, 
it stays perturbative to the GUT scale for $\lambda \leq 0.7$.
This upper bound is relaxed when one either takes the criteria 
that all couplings close to $\sim 4\pi$ for non-perturbation 
or allows new fields at the intermediate scale between the weak and GUT scale. 
We show that a hidden $U(1)_X$ gauge sector with spontaneously broken scale $\sim10$ TeV 
improves this bound as $\lambda\leq1.23$ instead.
This may induce significant effects on Higgs physics 
such as decreasing fine tuning involving the Higgs scalar mass, 
as well as on the small $\kappa$-phenomenology.

\end{abstract}

\maketitle

\section{I.~Introduction}
The standard model (SM)-like Higgs boson with mass around 126 GeV \cite{h} reported at the LHC strongly 
favors extension beyond the minimal supersymmetric model (MSSM).
A simple idea is adding a SM singlet $S$ to the MSSM, with superpotential 
\begin{eqnarray}{\label{NMSSM}}
W=\lambda~SH_{u}H_{d}+\frac{\kappa}{3}S^{3}.
\end{eqnarray}
Due to the Yukawa coupling $\lambda$ in Eq.(\ref{NMSSM}), 
the Higgs boson mass can be naturally uplifted to observed value
for moderate value $\lambda \geq 0.5$  at the electroweak (EW) scale.
This specific extension is  referred as 
the next-to-minimal supersymmetric model (NMSSM) \cite{0910.1785} for $\lambda\leq 0.7$, 
or $\lambda$-supersymmetry  ($\lambda$-SUSY) \cite{0607332} for $0.7\leq \lambda\leq 2$. 
Either  the NMSSM or $\lambda$-SUSY is very attractive from the viewpoint of naturalness argument \cite{1112.2703, 1312.0181},
as the stringent tension on stop masses required by the Higgs mass in the MSSM is dramatically reduced.
 
The origin of upper bound on $\lambda$  above can be seen from the beta function $\beta_{\lambda}$ for $\lambda$.
Since $\beta_{\lambda}$ is dominated by the top Yukawa coupling $y_t$,
the sign of which is positive,
it imposes an upper bound on the EW value of $\lambda$ 
in order to be still in the perturbative region at high energy scale.
Given this scale near the grand unification (GUT) scale $\simeq 1.0\times 10^{15}$ GeV, 
the critical value was found to be $\sim 0.7$ \cite{9801437} in terms of  the one-loop beta function
\footnote{In this paper, we follow the convention in \cite{9709356}, 
and present our $\beta$ functions in the $\overline{DR}$ scheme.},
\begin{eqnarray}{\label{nmssmbeta}}
\beta_{\lambda}=\frac{\lambda}{16\pi^{2}}\left[4\lambda^{2}+3y^{2}_{t}+2\kappa^{2
 }-3g^{2}_{2}-\frac{3}{5}g^{2}_{1}\right],
\end{eqnarray}
where $t\equiv\ln (\mu/\mu_{0})$, $\mu$ being the running renormalization scale and $\mu_{0}\equiv 1$ TeV.
On the other hand, 
the observed Higgs mass favors large value of $\lambda$ \cite{1310.0459}.

The intuition for solving the tension between the observed Higgs mass and perturbativity 
is obvious in the context of SUSY.
The main observation is that Yukawa couplings in superpotential receive only wave-function induced renormalizations due to the protection of SUSY,
in contrast to non-supersymmetric case \footnote{The phenomenon of Yukawa coupling being asymptotically free has been exposed in $\lambda\phi^{4}$ scalar field theory with dimension lower than four by Wilson and Fisher \cite{Wilson}, 
and in well-known nonlinear sigma model \cite{Polyakov} together with an attempt to obtain a version of four-dimensional supersymmetry \cite{Ketov}.   
Similar phenomenon was also addressed by authors of \cite{9406199} in four-dimensional scalar field theory with nonpolynomial potentials.}.
By following the fact that the beta functions of Yukawa couplings are related to the anomalous dimensions of chiral fields \cite{9709356},
and the anomalous dimensions are proportional to quadratic Yukawa couplings with positive coefficients,
the sign of contribution to $\beta$ function due to Yukawa interactions is thus always positive.
A way to alleviate this is introducing hidden super-confining gauge dynamics \cite{0405267, 0311349},
from which Yukawa couplings are asymptotically free, 
and $S$ is a composite other than fundamental state,
with large $\lambda$ at low energy as a result of this asymptotical freedom.
In this context the Higgs doublets can be either fundamental \cite{0405267} or composite states \cite{ 0311349}.

In this paper, we will explore $\lambda$-SUSY that stays perturbative up to the GUT scale in alternative way.
In this framework we will obtain a well defined $\lambda$-SUSY on the realm of perturbative analysis,
and operators such as Higgs doublets and singlet $S$  are all fundamental other than composite states.
This is our main motivation for this study.

The study of well defined $\lambda$-SUSY was initially addressed in Ref.\cite{9801437},
in which it was found that for $\tan\beta$ smaller than $\sim 10$,
$\lambda$ is the first Yukawa coupling running into the non-perturbative region at high energy scale.
It was also understood that either introducing new matter fields at the intermediate scale between the weak and GUT scale or adopting smaller EW value $\kappa(\mu_{0})$ can decrease the evolution rate for $\lambda$.
Given an initial EW value $\kappa(\mu_{0})$,
one can derive the upper bound on $\lambda(\mu_{0})$ or vice versa,
once new matters which appear at the intermediate scale are identified explicitly.
The minimal content of new fields only includes the messenger sector 
which communicates the SUSY breaking effect to the visible sector, 
namely the $\lambda$-SUSY.

Together with the messengers a spontaneously broken $U(1)_X$ gauge group 
will be considered as the new fields. 
We consider a class of models different from those studied in the literature \cite{0404251},  
in which SM fermions and sfermions except the Higgs doublets and $S$ are all charged under the hidden $U(1)_{X}$.
The abelian gauge coupling $g_{X}$ and the $U(1)_X$-breaking scale $M_{X}$ 
enter into the parameter space as two new free parameters.
In comparison with traditional NMSSM or $\lambda$-SUSY,
in our model the beta function for top Yukawa coupling $\beta_{t}$ receives 
new and negative contribution due to the hidden $U(1)_X$  sector,
the magnitude of which is determined by the hidden gauge coupling $g_{X}$ and scale $M_X$.
As long as $g_{X}$ is large enough but still valid for perturbative analysis, 
these new effects decrease the slope of $\lambda$ as function of renormalization scale $\mu$ above scale $M_{X}$,
and therefore lead to larger critical value $\lambda(\mu_{0})$.

The paper is organized as follows.
In section 2,  in terms of one-loop renormalization group equations (RGEs) for relevant coupling constants
we estimate the critical values of $\kappa(\mu_{0})$ and $\lambda(\mu_{0})$ 
due to the hidden $U(1)_X$ effect. 
In section 3, we revise the phenomenology for large $\lambda$ but small $\kappa$.
Finally we conclude in section 4.
In appendix A, we show the details of the hidden sector 
and briefly review collider constraints on the model parameters.

\section{II.~Perturbative $\lambda$-SUSY}
As mentioned in the introduction,
the beta function for Yukawa couplings of superpotential can be easily estimated by using the non-renormalization of superpotential.
Firstly, in the case without hidden $U(1)_{X}$ gauge group
the one-loop beta functions for Yukawa couplings in $\lambda$-SUSY are given by 
\footnote{The value of $\tan\beta$ defined as the ratio $\left<H^{0}_{u}\right>/\left<H^{0}_{d}\right>$ 
is required to be smaller than $\sim 10$ in light of the 126 Higgs mass in the NMSSM or $\lambda$-SUSY.
In \cite{0810.0989} it has been shown that experimental constraints require $\tan\beta \leq 10$ for $\lambda\geq1.0$.
Thus, it is a good approximation to ignore the bottom Yukawa coupling in compared with top Yukawa coupling.},
\begin{eqnarray}{\label{beta0}}
\beta_{\lambda}&=& \frac{\lambda}{16\pi^{2}} \left[ 4\lambda^{2}+2\kappa^{2}+3y^{2}_{t}-3g_{2}^{2}-\frac{3}{5}g^{2}_{1}\right],\nonumber\\
\beta_{y_{t}}&=& \frac{y_{t}}{16\pi^{2}} \left[ 6y^{2}_{t}+\lambda^{2}-\frac{16}{3}g^{2}_{3}-3g^{2}_{2}-\frac{13}{15}g^{2}_{1}\right] ,\nonumber\\
\beta_{\kappa}&=& \frac{\kappa}{16\pi^{2}} \left[6\lambda^{2}+6\kappa^{2} \right].
\end{eqnarray}
The one-loop beta functions for the SM gauge couplings are the same as the MSSM,
\begin{eqnarray}{\label{gauge}}
\frac{\partial}{\partial t} \alpha^{-1}_{i} &=& -\frac{b_{i}}{2\pi}.
\end{eqnarray}
where $(b_{1}, b_{2}, b_{3})=(33/5, 1, -3)$.

In the case with a hidden $U(1)_{X}$ gauge group, 
we study the model in which SM fermions and sfermions all carry a hidden $U(1)_{X}$ charge, 
with the spontaneously broken scale $M_{X}\simeq 10$ TeV. 
The Higgs doublets, however, are singlets of this $U(1)_{X}$ symmetry.
Anomaly free conditions require three hidden matters $X_{1,2,3}$ with the same $U(1)_{X}$ charge added to the model.
For details about the matter representations, see appendix A.
The modifications to RGEs in Eq.(\ref{beta0}) due to the hidden $U(1)_{X}$ effect are given by,
\begin{eqnarray}{\label{beta}}
\delta \beta_{\lambda}&=&0,\nonumber\\
\delta \beta_{y_{t}}&=&-\frac{y_{t}}{16\pi^{2}}\left(\frac{12}{5}a^{2}\cdot g^{2}_{X}\right), \nonumber\\
\delta \beta_{\kappa}&=&0,
\end{eqnarray}
together with 
\begin{eqnarray}{\label{gx}}
\beta_{g_{X}}=\frac{g^{3}_{X}}{16\pi^{2}}\left[\frac{3}{5}a^{2}\cdot\left(\frac{39}{2}n_{g}+3\right)\right],
\end{eqnarray}
where $n_{g}=3$ denotes the number of SM fermion generations.
It is obvious that large EW value $g_{X}(\mu_{0})$ favors small $\mid a\mid$.

The RG running for $g_X$ with different values at $\mu_{0}$ is plotted in Fig.1.
In this figure non-perturbative region lies below the red line.
It shows the critical value $g_{X}(\mu_{0})\leq \{2.0, 1.18, 0.82\}$ for $a=\{\frac{1}{6},\frac{1}{3},\frac{1}{2}\}$, respectively.
Given the upper bound on $g_{X}(\mu_{0})$, 
the upper bound on Yukawa coupling $\lambda$ can be estimated 
in terms of RGEs in Eq.(\ref{beta}) and Eq.(\ref{beta0}).
The input parameters related to the critical value $\lambda(\mu_{0})$ are composed of
\begin{eqnarray}{\label{input}}
\{y_{t}(\mu_{0}),~\kappa(\mu_{0}),~g_{X}(\mu_{0}),~M,~n_{5\bar{5}}\},
\end{eqnarray}
where $M$ denotes the messenger mass scale, 
$n_{5\bar{5}}$ represents the number of $\bar{5}+5$ representation of $SU(5)$ for messengers.
We will use the updated pole mass of top quark $m_{t}=174$ GeV \cite{top} 
instead of $m_{t}=180$ GeV in \cite{9801437} for our analysis,  
and take the criteria that the theory enters into the non-perturbative region 
when any of coupling constants in the theory is bigger than $\sim 4\pi$.
\begin{center}
\begin{figure}
\includegraphics[width=0.45\textwidth]{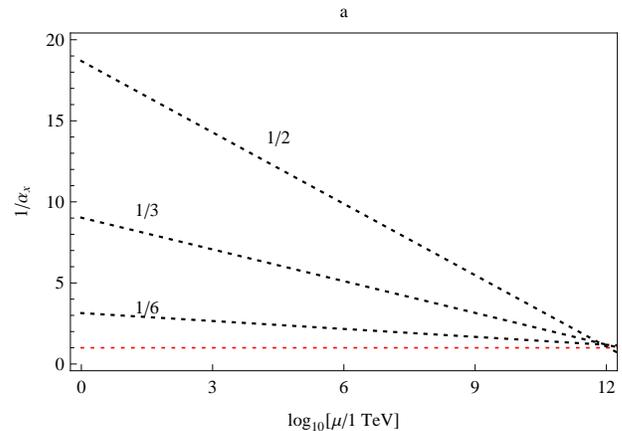}
\caption{Critical value $g_{X}(\mu_{0})$ for $a=\{\frac{1}{6},\frac{1}{3},\frac{1}{2}\}$, respectively.}
\end{figure}
\end{center}

We want to emphasize that $(i)$, below scale $M_{X}$, 
the modifications to RGEs arising from $U(1)_{X}$ can be ignored. 
$(ii)$, above scale $M$, 
coefficients $b_{i}$ in beta functions for SM gauge couplings should change as $b_{i}\rightarrow b_{i}+n_{5\bar{5}}$.
Fig. 2 shows the modifications to the upper bound on $\kappa(\mu_{0})$ due to the hidden $U(1)_{X}$ sector 
given fixed $\lambda(\mu_{0})$.
We choose the initial EW value $\lambda(\mu_{0})=1.0$,  $g_{X}(\mu_{0})=0.82$ for $a=1/2$, 
and messenger paramaters $M=10^{7}$ GeV and $n_{5\bar{5}}=\{1,4\}$ for illustration.
Given the critical value $\kappa(\mu_{0})$, 
the solid line in gray (green) represents the RG running for $\lambda^{-1}$ without (with) $U(1)_X$ effect.
The RG runnings of $\kappa^{-1}$ (dotted) and $y^{-1}_{t}$ (dotdashed) shown in the figure imply that the upper bound $\kappa_{0}\leq 0.8$ is improved to $\kappa_{0}\leq 0.85\sim 0.86$ when the hidden $U(1)_{X}$ effect is taken into account.
Alternatively, given the same $\kappa(\mu_{0})$, the upper bound on $\lambda(\mu_{0})$ can be improved by the $U(1)_{X}$ effect. 
Combination of the left and right panel shows that the deviation due to the change of $n_{5\bar{5}}$ is small, in comparison with the $U(1)_X$ effect.
Note that plots in Fig.1 are actually critical lines, 
because the change from the perturbative to the non-perturbative region is rather abrupt.

\begin{widetext}
\begin{center}
\begin{figure}[ht]
\centering
\begin{minipage}[b]{0.5\textwidth}
\centering
\includegraphics[width=3in]{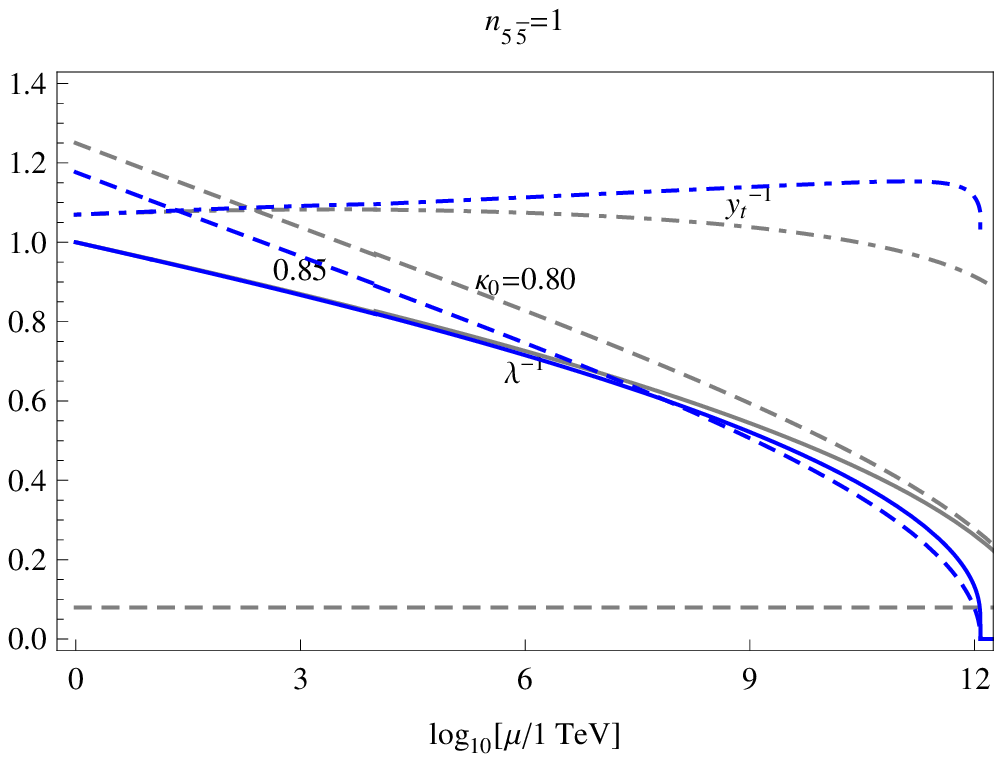}
\end{minipage}%
\centering
\begin{minipage}[b]{0.5\textwidth}
\centering
\includegraphics[width=3in]{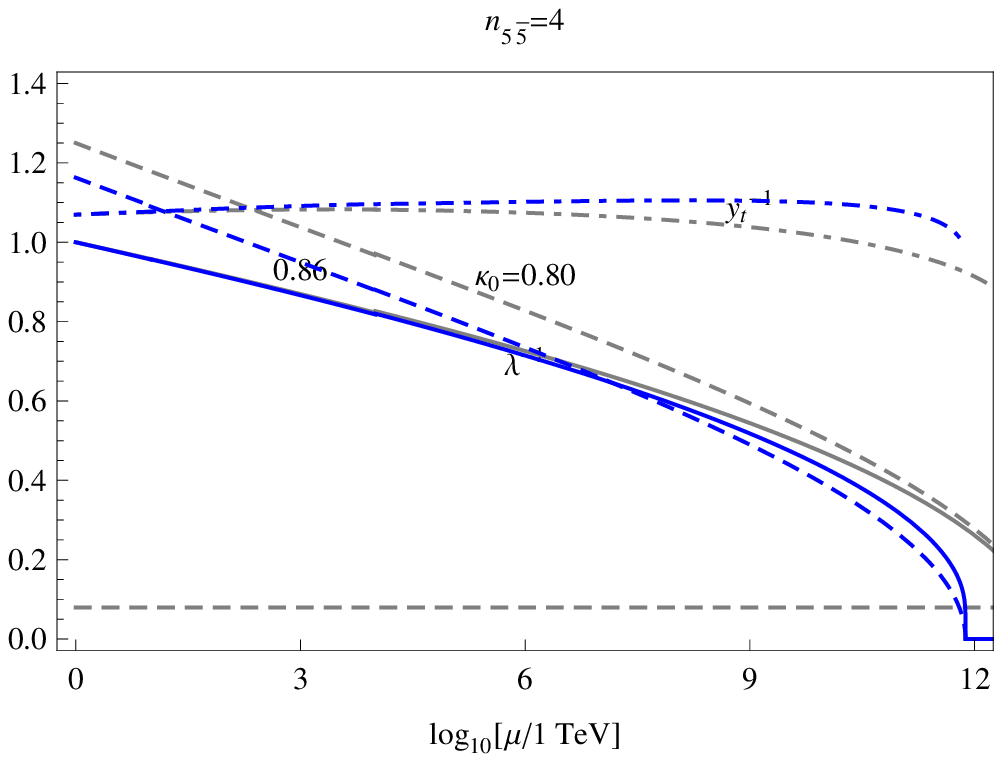}
\end{minipage}%
\caption{RG running of $\lambda^{-1}$ (solid line) for initial EW value $\lambda(\mu_{0})=1.0$ and $g_{X}(\mu_{0})=0.82$ for $a=1/2$,  and messenger parameters $M=10^{7}$ GeV and $n_{5\bar{5}}=1$ (left) and $n_{5\bar{5}}=4$ (right), respectively.
The gray and green color corresponds to $\lambda$-SUSY without and with $U(1)_{X}$, respectively.
The RG runnings of $\kappa^{-1}$ (dotted) and $y^{-1}_{t}$ (dotdashed) shown in the figure indicate that the upper bound $\kappa_{0}\leq 0.8$ is improved to $\kappa_{0}\leq 0.85\sim 0.86$ when the hidden $U(1)_{X}$ effect is taken into account.
Horizontal line refers to critical point between perturbative and non-perturbative region.  }
\end{figure}
\end{center}
\end{widetext}

 \begin{table}
\begin{center}
\begin{tabular}{|c|c|c|}
  \hline
   $\lambda$(1TeV) & \ without $U(1)_X$ & \ with $U(1)_X$ \\
  \hline\hline
  1.0 & $\kappa(\mu_{0})\leq 0.80$ & $\kappa(\mu_{0})\leq 0.85$ \\
  1.10 & $\kappa(\mu_{0})= 0$ & $\kappa(\mu_{0})\leq 0.81$\\
  1.23&  not well defined & $\kappa(\mu_{0})= 0$ \\
  \hline
\end{tabular}
\caption{Upper bound on $\lambda(\mu_{0})$ for $g_{X}(\mu_{0})=0.82$ for $a=1/2$, 
 messenger parameters $M=10^7$ GeV and $n_{5\bar{5}}=1$, and different initial values of $\kappa(\mu_{0})$.
In contrast to the result $\lambda(\mu_{0})\leq 0.8$ for $\kappa(\mu_{0})\simeq 0$ in \cite{9801437},
we have $\lambda(\mu_{0})\leq 1.23$ instead for the case with the hidden $U(1)_{X}$.}
\end{center}
\end{table}

In table I, we show the improvement on the upper bound on $\lambda(\mu_{0})$ 
due to the hidden $U(1)_{X}$. 
Without $U(1)_X$ sector, the model is not well defined up to the GUT scale when $\lambda >0.80$.
Instead, $\lambda\leq 1.23$ stays perturbative up to the GUT scale when the hidden $U(1)_X$ is added to the model.
The critical value on $\lambda$ may be modified due to different choices on parameters $\{n_{5\bar{5}}, M, a\}$.
As shown in Fig.2,  the deviation to $\lambda(\mu_{0})$ due to messenger parameters $M$ and $n_{5\bar{5}}$ is rather small.
Similar conclusion holds when one tunes $a$.
Because $a$ larger than $1/2$ leads to larger contribution to $\delta\beta_{y_{t}}$ but 
unfortunately smaller $g_{X}(\mu_{0})$, and vice versa.

In summary, in terms of introducing a hidden $U(1)_{X}$ sector with gauge
symmetry broken scale $\simeq 10$ TeV 
well defined $\lambda$-SUSY up to the GUT scale is allowed for $\lambda\leq 1.23$.
In this class of models all fields including singlet $S$ and the Higgs doublets are fundamental. 
The implication for $\lambda\sim 1-2$ has been addressed, e.g., in \cite{1310.0459}.
In the next section, we revise the phenomenological implication for such  large $\lambda$ together with small $\kappa$.

\section{III.~PQ Symmetry and Small $\kappa$-Phenomenology}
This section is devoted to study the phenomenology in $\lambda$-SUSY with small $\kappa$.
Although it is a consequence of taking large $\lambda$ for well defined $\lambda$-SUSY,
the study of small $\kappa$-phenomenology can be considered as an independent subject from the viewpoint of phenomenology.
The smallness of $\kappa$ can be understood due to a broken $U(1)$ global symmetry, 
i.e., Peccei-Quinn (PQ) symmetry. 
Without $\kappa$, the model is invariant under the following $U(1)$ symmetry transformation,
\begin{eqnarray}{\label{PQ}}
H_{u}\rightarrow H_{u} \exp({i\phi}), ~H_{d}\rightarrow H_{d} \exp({i\phi}), ~
S\rightarrow S \exp({-2i\phi}), \nonumber\\
\end{eqnarray}
Terms like $\delta V=m^{2}S^{2}+B_{\mu} H_{u}H_{d}$ explicitly breaks this symmetry.
For these breaking small, we have a pseudo-Goldstone boson with small mass.

The soft terms in the scalar potential for small $\kappa$-phenomenology is given by,
\begin{eqnarray}{\label{soft}}
V_{soft}&=&m^{2}_{S}\mid S \mid ^{2}+m^{2}_{H_{u}} H^{2}_{u}+m^{2}_{H_{d}} H^{2}_{d}\nonumber\\
&+&(A_{\lambda}\lambda SH_{u}H_{d}+ \text{H.c}),
\end{eqnarray}
where $\kappa$ relevant terms are ignored.
The EW symmetry breaking vacuum can be determined from Eq.(\ref{soft}).
For details on analysis of EW breaking vacuum, 
see, e.g., \cite{0910.1785, 0712.2903}.
There are five free parameters
\begin{eqnarray}
\{m^{2}_{H_{u}},~m^{2}_{H_{d}},m^{2}_{S}, ~A_{\lambda},~\lambda\}
\end{eqnarray}
which define the small $\kappa$-phenomenology.
The first two can be traded for $\upsilon=174$ GeV and $\tan\beta$.
In what follows we explore the constraints on these parameters,
the Higgs scalar spectrum, and their couplings to SM particles.

The $3\times3$ squared mass matrix of CP-even neutral scalars reads 
\footnote{Here we follow the conventions and notation in Ref.\cite{0712.2903}. },
\begin{widetext}
\begin{eqnarray}{\label{matrix}}
\mathcal{M}^{2}=\left(%
\begin{array}{ccccc}
  \frac{A_{\lambda}^{2}}{1+x}+(M^{2}_{Z}-\lambda^{2}\upsilon^{2})\sin^{2}2\beta  & -\frac{1}{2}(M^{2}_{Z}-\lambda^{2}\upsilon^{2}) \sin4\beta& - A_{\lambda}\lambda\upsilon\cos 2\beta  \\
 * & M^{2}_{Z}\cos^{2}2\beta+\lambda^{2}\upsilon^{2}\sin^{2}2\beta & - A_{\lambda}\lambda\upsilon\sin 2\beta \frac{x}{1+x}  \\
* & *& \lambda^{2}\upsilon^{2}(1+x)   \\
\end{array}%
\right)
\end{eqnarray}
\end{widetext}
in the basis $(H, h, s)$,
where $x\equiv m^{2}_{S}/(\lambda\upsilon)^{2}$,
 $H=\cos\beta h_{2}-\sin\beta h_{1}$ and $h=\cos\beta h_{1}+\sin\beta h_{2}$
under the compositions $H_{u}^{0}=\frac{1}{\sqrt{2}}(h_{1}+i\pi_{1})$, 
$H_{d}^{0}=\frac{1}{\sqrt{2}}(h_{2}+i\pi_{2})$ and $S=\frac{1}{\sqrt{2}}(s+i\pi_{s})$.
The state $h$ mixes with other two states $H$ and $s$ via $\mathcal{M}^{2}_{12}$ and $\mathcal{M}^{2}_{23}$ , respectively.
Due to the smallness of $\mathcal{M}^{2}_{12}$,
state $h$ mixes dominately with $s$.
Without such mixing ( i.e., $\mathcal{M}^{2}_{23}\rightarrow 0$ ),
$h$ couples to SM gauge bosons and fermions exactly as the SM Higgs.
This implies that in the parameter space of small $m_{S}$,
$h$ is the SM-like scalar discovered at the LHC with $\mathcal{M}^{2}_{22}=(126 \ GeV)^{2}$,
and $s$ decouples from SM gauge bosons with mass $\lambda \upsilon\sim 209$ GeV for $\lambda=1.2$ .
Such scalar $s$ easily escapes searches performed by the LEP2 and Run-I of LHC. 

The precision measurement of $h$ including its couplings to SM gauge bosons and fermions
powerfully constrains the magnitude of mixing effect.
After mixing,  we define the mass eigenstates as $h_{1}$ and $h_{2}$,
where $h_{1}=\cos \theta h-\sin\theta s$, $h_{2}=\cos \theta s +\sin\theta h$.
Normalized to SM Higgs boson couplings,
$h_1$ and $h_2$ couple to SM particles as,
 \begin{eqnarray}{\label{coupling}}
\xi_{h_{1}VV}&=&\frac{g^{2}_{h_{1}VV}}{g^{2}_{hVV}}=\cos^{2}\theta,~~
\xi_{h_{2}VV}=\frac{g^{2}_{h_{2}VV}}{g^{2}_{hVV}}=\sin^{2}\theta.\nonumber\\
\end{eqnarray}
where $V=\{W, Z, t, b,\cdots \}$.
At present status, LHC data suggests that $0.96\leq\cos^{2}\theta\leq 1$ \cite{fit}. 
Alternatively we have $\sin^{2}\theta \leq 0.04$.

In Fig.3 we show the parameter space in the two-parameter plane of $m_S$ and $A_{\lambda}$,
for $\lambda=1.2$ and $\tan\beta=\{4.5, 5, 5.5\}$, respectively.
For each $\tan\beta$ region below the color contour is excluded by the condition of stability of potential
and mass bound on chargino mass $m_{\chi}>103$ GeV \cite{chargino}. 
Region in the right up corner is excluded by the precision measurement of Higgs coupling presently.
In particular, $m_{S}$ heavier than $\sim$90, 105, and 110 GeV is excluded for $\tan\beta=4.5$, 5 and 5.5, respectively.
In comparison with the choice $\lambda=0.7$  discussed in \cite{0712.2903},
the main difference is that for $\lambda=1.2$ it allows larger $m_{S}$. 

As for the Higgs mass constraint,
the discrepancy between $\mathcal{M}^{2}_{22}$ and 126 GeV is compensated by the stop induced loop correction. 
The stop mass for the zero mixing effect (i.e., $A_t\sim 0$) is $\sim 340$ GeV for $\tan\beta=4.5$, $\sim 550$ GeV for $\tan\beta=5$ and $\sim 800$ GeV for $\tan\beta=5.5$. 
Stop mass beneath 1 TeV is favored by naturalness.
Nevertheless, there has tension for such light stop with present LHC data.
This problem can be resolved in some situation,
which we will not discuss here.

We show in Fig.4 the ratio $\xi_{h_{2}VV}$ defined in Eq.(\ref{coupling}),
which determines the production rate for scalar $h_2$.
Its magnitude increases slowly as $m_S$ becomes larger.
$\xi_{h_{2}VV}$ reaches $\sim 0.03$ at most when $m_S$ closes to its upper bound (suggested by Fig. 3).
For such strength of coupling and mass $\sim 200$ GeV,
$h_2$ is easily out of reach of Run-I at the LHC and earlier attempts at the LEP2.
Small ratio of strength of coupling similarly holds for heaviest CP-even neutral state $H$. 
In this sense, it is probably more efficient to probe charged Higgs scalar $H^{\pm}$, CP-odd scalar $A$, or light pseudo-boson $G$.
Studies along this line can be found in, e.g, \cite{0005308}.
\begin{figure}[ht]
\includegraphics[width=0.45\textwidth]{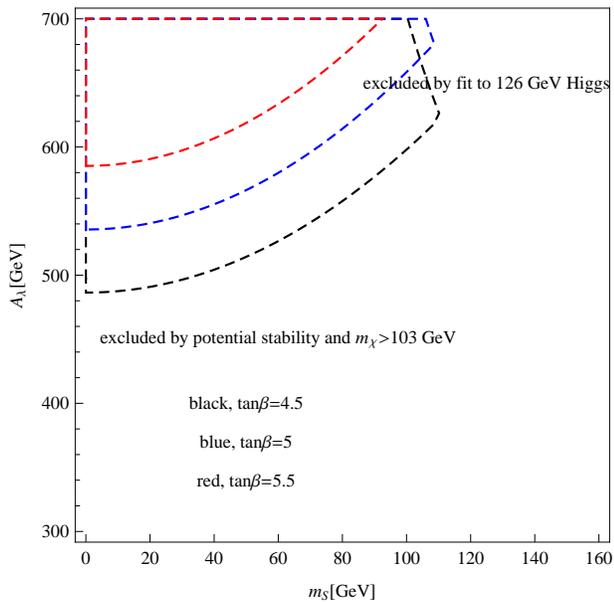}
\caption{Parameter space in the two-parameter plane of $m_S$ and $A_{\lambda}$,
for $\lambda=1.2$ and $\tan\beta=4.5$ (black), 5 (blue), 5.5 (red), respectively.
For each $\tan\beta$ region below the color contour is excluded by the condition of stability of potential
and mass bound on chargino mass $m_{\chi}>103$ GeV. 
Region in the right up corner is excluded by the fit to Higgs couplings presently.}
\end{figure}

\begin{figure}
\includegraphics[width=0.45\textwidth]{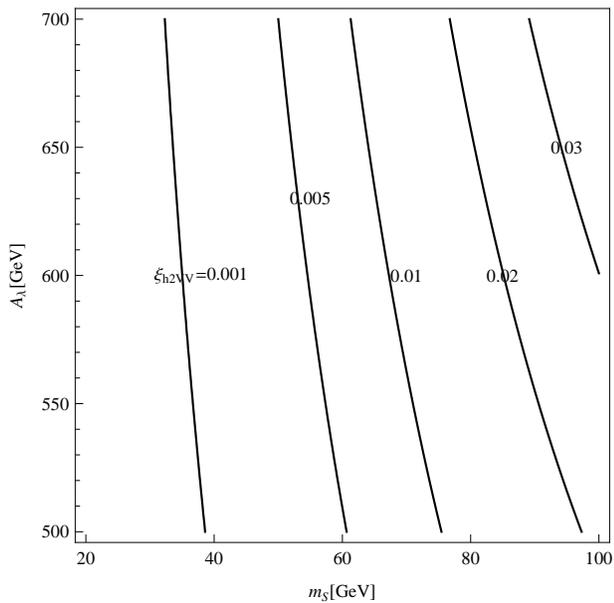}
\caption{Ratio of coupling strength for $h_2$ shown in the plane of $m_{S}-A_{\lambda}$ for $\lambda=1.2$ and $\tan\beta=4.5$.}
\end{figure}

As for other scalar masses we show them in table II for  three sets of $A_{\lambda}$ and $m_{S}$,
which are chosen in the parameter space shown in Fig.3.
It is shown that for each case the mass of charged Higgs boson exceeds its experimental bound $\sim 300$ GeV,
and $A$ scalar is always the heaviest with mass around 600 GeV.
In comparison with scalar mass spectrum in fat Higgs model \cite{0311349},
the spectrum in table II is similar to it,
although their high energy completions are rather different.
As a final note we want to mention that the smallness of $m_S$ in compared with $A_{\lambda}$ can be achieved in model building, e.g., in gauge mediation. 
Because singlet $S$ only couples to messengers indirectly through Higgs doublets. 
\begin{table}
\begin{center} 
\begin{tabular}{|c|c|c|c|}
  \hline
   $(A_{\lambda}, m_{S})$ (GeV) & $H^{\pm}$ (GeV) & $H$ (GeV)& $A$(GeV)\\
  \hline\hline
  (500,~40) & 463& 482& 535 \\
 (560,~50) & 527 & 530 & 585 \\
 (620,~50) & 590 & 586 & 640 \\
  \hline
\end{tabular}
\caption{Mass spectrum for three sets of choices, which are subtracted from Fig.3.}
\end{center}
\end{table}

\section{IV.~Conclusions}
In this paper, we have studied $\lambda$-SUSY which stays perturbative up to the GUT scale.
We find that the bound $\lambda\leq0.7\sim 0.8$ in the minimal model
is relaxed to $\lambda\leq1.23$ 
if a hidden $U(1)_X$ gauge theory is introduced 
above scale $\sim 10$ TeV 
and small $\kappa(\mu_{0})$ is assumed at the same time.
This improvement gives rise to several interesting consequences in phenomenology.
For example, the fine tuning related to 126 Higgs mass can be reduced,
and light stop beneath 1 TeV can be allowed.
In the light  of such single $U(1)_X$,
one may introduce multiple $U(1)$s or other gauge sectors at the intermediate scale,
and further uplift the bound on $\lambda$.

In the second part of the paper, 
we have revised small $\kappa$-phenomenology for large $\lambda$.
In comparison with fat Higgs model \cite{0311349},
the spectrum in the small $\kappa$-phenomenology is similar,
although their high energy completions are different.
The null result for signals of the other two CP-even neutral scalars $h_2$ and $H$
is due to the perfect match between the scalar 
discovered at the LHC (Here its is referred as $h_1$) and the SM Higgs.
Because the perfect fit dramatically reduces the mixing effect between $h_1$ and the others,
which results in tiny strength of coupling for $h_2$ and $H$ relative to the SM expectation.
The studies on signals of charged scalar $H^{\pm}$, CP-odd scalar $A$ and pseudo-boson $G$
will shed light on this type of model.\\

~~~~~~~~~~~~~~~~~~~
$\mathbf{Acknowledgement}$\\
The author thanks J-J. Cao for correspondence and the referee for valuable suggestions.
This work is supported in part by Natural Science Foundation of China under Grant No.11247031 and 11405015.

\appendix
\section{A. The Hidden $U(1)_X$ sector}
\begin{table}
\begin{center}
\begin{tabular}{|c|c|c|}
  \hline\hline
   & $SU(3)\times SU(2)\times U(1)_{Y}$ & $U(1)_{X}$ \\
  \hline\hline
  $Q_{i}$ & $(\mathbf{3},\mathbf{2}, \frac{1}{6})$ & $a$ \\
  $\bar{u}_{i}$ & $(\bar{\mathbf{3}},\mathbf{1}, -\frac{2}{3})$ & $-a$ \\
  $\bar{d}_{i}$ & $(\bar{\mathbf{3}}, \mathbf{1}, \frac{1}{3})$ & $-a$ \\
  $L_{i}$ & $(\mathbf{1}, \mathbf{2}, -\frac{1}{2})$ & $-3a$ \\
  $\bar{e}_{i}$ & $(\mathbf{1},\mathbf{1}, 1)$ & $3a$ \\
  $H_{u}$ & $(\mathbf{1}, \mathbf{2}, \frac{1}{2})$ & $0$ \\
  $H_{d}$ & $(\mathbf{1}, \mathbf{2}, -\frac{1}{2})$ & $0$ \\
 \hline
  $X_{1,2,3}$ & $(\mathbf{1},\mathbf{1}, 0)$ & $3a$ \\
  $S$ & $(\mathbf{1},\mathbf{1}, 0)$ & $0$ \\
  \hline
\end{tabular}
\caption{Charge assignments for the matter superfileds in the visible and hidden sectors.
$Q_{i}$, $\bar{u}_{i}$, $\bar{d}_{i}$, $L_{i}$ and  $\bar{e}_{i}$ denote the three-generation matters of MSSM.
$S$ is a singlet of both SM gauge group and the hidden $U(1)_X$.
$X_{1,2,3}$ are a set of hidden matters charged under the hidden $U(1)_X$,
which are responsible for the $U(1)_X$ gauge symmetry breaking.
$a$ is a real number.}
\end{center}
\end{table}

$\mathbf{Realistic~Model}$.
The model that we are going to study for the case with hidden $U(1)_X$ gauge symmetry is presented in table III.
The $U(1)_X$ charges in this table must satisfy the gauge anomaly free conditions and 
are consistent with the superpotential of visible sector, $W\sim y_{u}Q_{i}\bar{u}_{i}H_{u}+y_{d}Q_{i}\bar{d}_{i}H_{d}+y_{e}L_{i}\bar{e}_{i}H_{d}+\mu H_{u}H_{d}$.
We limit to the case in which the Higgs doublets $H_{u,d}$ are singlets of $U(1)_{X}$.
Without the three hidden matters $X_{1,2,3}$ with the same $U(1)_{X}$ charge 
anomaly free conditions such as $U(1)_{X}-Graviton-Graviton$ and $U(1)_{X}-U(1)_{X}-U(1)_{X}$ can not be satisfied \footnote{
We thank the referee for reminding us that the $U(1)_{X}$ charges in this model are identical to $B-L$ quantum numbers of the SM fields, 
and the charges of the "hidden" matter, $X_{1,2,3}$ are same as those of right-handed neutrino fields.}.

$\mathbf{Symmetry~Breaking}$.
There is no signal of $Z'$ from broken $U(1)_X$ gauge group yet,
so it should be spontaneously broken above the weak scale for $g_{X}$ of order SM gauge coupling.
This can be achieved in various ways.
Here we simply take the gauge mediation for example.
If the hidden $U(1)_{X}$ gauge group communicates the D-type of SUSY-breaking effects into $X$s,
the sign of soft mass squared $m^{2}_{X}$ would be negative \cite{0705.0865}.
For earlier application of such property in gauge mediation, see, e.g, \cite{0812.4600}. 
Below SUSY-breaking scale the potential for $X_i$ we have
\begin{eqnarray}{\label{V}}
V=-m^{2}_{X_{i}} X_{i}^{\dag}X_{i}+ \left(X_{i}^{\dag}X_{i}\right)^{2}.
\end{eqnarray}
which spontaneously breaks $U(1)_X$, with $U(1)_X$-breaking scale $M_{X}\sim m_{X}$.
Note that the magnitude of $m_{X}$ can be either larger or smaller than the soft breaking masses in the visible sector,
which depends on the ratio of $D$ term relative to $F$ term and also the ratio of $g_{X}$ relative to SM gauge couplings.
With a $D$ term which gives rise to an order of magnitude larger than soft breaking masses
and $g_{X}$ of the same order as SM gauge coupling,
one can obtain $M_{X}\simeq \mathcal{O} (10)$ TeV and $m_{\tilde{f_{i}}}\sim \mathcal{O} (1)$ TeV in the visible sector.

$\mathbf{Limit~ on~ g_{X}(\mu_{0})}$.
The experimental constraint on $M_{X}$ and gauge coupling $g_{X}$ 
is obtained from direct production of $Z'$ at colliders through leptonic decays;
and also indirect searches from flavor violation and electroweak precision tests.
For a review on the status of hidden $U(1)_X$, see e.g., \cite{Zprime}.
The parameter $M_X$ is constrained to be above $1\sim 2$ TeV for coupling $g_{X}\sim 1-2$.
For $M_{X}\simeq 10$ TeV adopted in this note,  we have the experimental limit $g_{X}\leq M_{X}/(\mid a\mid \cdot \text{a few TeV})\sim \mid a\mid^{-1}$  \cite{limit}.

\end{document}